\documentclass[11pt]{article}
 \usepackage[hmargin=1.25in,vmargin=1in]{geometry}
 \usepackage{amsmath}
 \usepackage{graphicx}
 \usepackage{subfig}
 \usepackage{booktabs}
 \usepackage{cite}
 \usepackage{hyperref}
 \usepackage{tabularx}
 \usepackage{ragged2e}
 \usepackage{booktabs} 

\title{Recognition of Cardiac MRI Orientation via Deep Neural Networks and a Method to Improve Prediction Accuracy}

\author{Houxin Zhou}

\begin{document}

\maketitle
\bibliographystyle{plain}

\begin{abstract}
 	In most medical image processing tasks, the orientation of an image would affect  computing result. However, manually reorienting images wastes time and effort. In this paper, we study the problem of recognizing orientation in cardiac MRI and using deep neural network to solve this problem. For multiple sequences and modalities of MRI, we propose a transfer learning strategy, which adapts our proposed model from a single modality to multiple modalities. We also propose a prediction method that uses voting. The results shows that deep neural network is an effective way in recognition of cardiac MRI orientation and the voting prediction method could improve accuracy.
\end{abstract}
\section{Introduction}

When medical images were stored, they may have different image orientations. In the further segmentation or computing, this difference may affect the results, since current deep neural network (DNN) systems generally only take the input and output of images as matrices or tensors, without considering the imaging orientation and real world coordinate. So it is crucial to recognize it before the further computing. This work is aimed to provide a study of the Cardiac Magnetic Resonance (CMR) image orientation, for fitting the coordinate system of human reality, and to develop an efficient method for recognition of the orientation.

Deep neural network has performed outstandingly in computer vision and gradually replaced the traditional methods. DNN also take a important role in medical image processing, such as image segmentation\cite{R:1} and myocardial pathology analysis\cite{R:2}. For CMR images, standardization of all the images is a prerequisite for further computing tasks based on DNN-based methodologies.

Most studies in the field of medical image processing have only focused on the further computing, so they have to spend a lot of manpower to do the preprocess. If we can auto adjust the images, it will save lots of time. Nevertheless, recognizing the orientation of different modality CMR images and adjusting them into standard format could be as challenging as the further computing tasks\cite{R:3}. In a broad sense, recognition orientation is also a kind of image classification task, so DNN is of no doubts an effective way to solve this problem. In this work, we still use DNN as our main method.

In most image classification problem like ImageNet, we do some transformation to the image but these transformation do not change the label, for example we rotate a dog image and it's still a dog image. However, the orientation could be changed if we do transformation like flipping to the images. In this work, we utilize this character to built a predicting model. Combining DNN and predicting model, we built a framework for recognition of Orientation.

This work is aimed at designing a DNN-based approach to achieve orientation recognition for multiple CMR modalities. Figure \ref{fig:1} presents the pipeline of our proposed method. The main contributions of this work are summarized as follows:

\begin{figure}[htb]
\centering 
\includegraphics[width=1.0\linewidth]{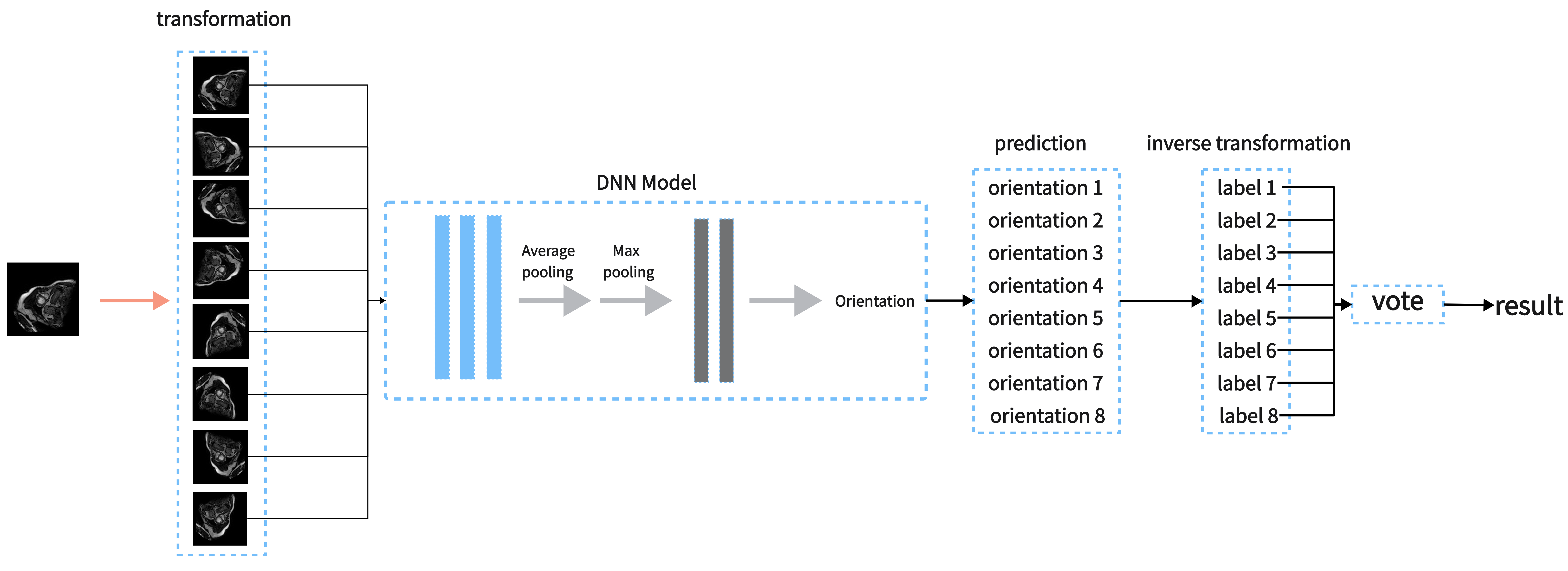}
\caption{The pipeline of our proposed method. Firstly, we do some transformatioin to the image(See 2.1 and 2.3). Then, the images are taken as input to the DNN model and generate orientations. Finally, we apply inverse transformation to these orientations and vote to get the result.}
\label{fig:1}
\end{figure}

\begin{enumerate}
	\item We propose a scheme to standardize the CMR image orientation and categorize all the orientations for classification.
	\item We present a DNN-based orientation recognition method for CMR image and transfer it to other modalities.
	\item We propose a predicting method to improve the accuracy for orientation recognition.
\end{enumerate}

\section{Method}
In this section, we introduce our proposed method for orientation recognition. Our proposed method is based on Deep Neural Network which was proved effective in image classification. In CMR Image Orientation Categorization, we improved the predicting accuracy by the following four steps. Firstly, we apply invertible operators to the image to get another 7 images. Then we predict these images and get 8 orientations. Finally, we use inverse transformation to these orientations and then vote to get the result.

\subsection{CMR Image Orientation Categorization}
Due to different data sources and scanning habits, the orientation of different CMR images may be different, and the orientation vector corresponding to the image itself may not correspond correctly. This may cause problems in tasks such as image segmentation or registration. Taking a 2D image as an example, we set the orientation of an image as the initial image and set the four corners of the image to $\boxed{\begin{array}{cc}
	1 & 2\\
	3 & 4 \\
\end{array}}$ , Then the orientation of the 2D MR image may have the following 8 variations, which is listed in Table \ref{Table 1}.
	
	For each image $X_t$ from dataset, the target is to find the correct orientation from 8 classes. We denote the correct orientation of image $X_t$ as $i_t$ and denote the correctly adjusted image $X_t$ as $Y_t$ If we view each orientation as a function $f$, we can get a function set $\{f_i, i = 0\dots 7\}$ and $(X_t, Y_t,i_t)$ satisfy the equation $f_{i_t}(Y_t) = X_t$. In the following, function set $\{f_i, i = 0\dots 7\}$ is referred to as $F$.
	
\begin{table}[htb]
	\centering
	\caption{Orientation categorization of 2D CMR Images. Here, sx, sy and sz respectively denote the size of image in X-axis, Y-axis and Z-axis.}
	\label{Table 1}
	\resizebox{1.0\columnwidth}{!}{
	\begin{tabular}{|l|lll|}
	\hline
	No.&Operation&Image&Correspondence of coordinates\\
	\hline
		0 & initial state & $\boxed{\begin{array}{cc}
	1 & 2\\
	3 & 4 \\
\end{array}}$ &Target[x,y,z]=Source[x,y,z]\\
	\hline
		1&horizontal flip&$\boxed{\begin{array}{cc}
	2 & 1\\
	4 & 3 \\
\end{array}}$ &Target[x,y,z]=Source[sx-x,y,z]\\
	\hline
		2&vertical flip&$\boxed{\begin{array}{cc}
	3 & 4\\
	1 & 2 \\
\end{array}}$ &Target[x,y,z]=Source[x,sy-y,z]\\
	\hline
		3&Rotate $180^{\circ}$ clockwise&$\boxed{\begin{array}{cc}
	4 & 3\\
	2 & 1 \\
\end{array}}$ &Target[x,y,z]=Source[sx-x,sy-y,z]\\
	\hline
		4&Flip along the upper left-lower right corner&$\boxed{\begin{array}{cc}
	1 & 3\\
	2 & 4 \\
\end{array}}$ &Target[x,y,z]=Source[y,x,z]\\
	\hline
		5&Rotate $90^{\circ}$ clockwise&$\boxed{\begin{array}{cc}
	3 & 1\\
	4 & 2 \\
\end{array}}$ &Target[x,y,z]=Source[sx-y,x,z]\\
	\hline
		6&Rotate $270^{\circ}$ clockwise&$\boxed{\begin{array}{cc}
	2 & 4\\
	1 & 3 \\
\end{array}}$ &Target[x,y,z]=Source[y,sy-x,z]\\
	\hline
		7&Flip along the bottom left-top right corner&$\boxed{\begin{array}{cc}
	4 & 2\\
	3 & 1 \\
\end{array}}$ &Target[x,y,z]=Source[sx-y,sy-x,z]\\
	\hline
	\end{tabular}}
	\end{table}
	\subsection{Deep Neural Network}
	Suppose given image $X_t$, $X_t$ is then normalized. We denote the processed $X_t$ as $X_t'$. CNN take the $(X_t',i_t)$ as input. In the proposed framework, the orientation recognition network consists of 3 convolution layers and 2 fully connected layers. The orientation predicted is denoted as $\hat{O_i}$. We use the standard categorical loss to calculate the loss between predicted orientation $\hat{O_i}$ and orientation label $O_i$. The orientation loss is formulated as below:
	\begin{equation}
		L_{orientation} = \sum_{i = 1}^8O_ilog(\hat{O_i})
	\end{equation}
	\subsection{Improved Prediction Method}
	As we can see in 2.1, we regard label $i_t$ as function $f_{i_{t}}$. It can be easily proved that $f_i \circ f_j \in F$ for any $i$ and $j$, so we can not only view $f_i$ as a function but an operator whose define domain and value domain are both $F$. For convenience, we denote the operator $f_i$ as $g_i$. Surpprisingly, we can prove that operator $g_i$ is a surjection in $F$ for any $i$, which can be simply explained by the following matrix $A$. In matrix $A$, $A_{ij} = k$ means $g_i(f_j) = f_k$.
	\[A = \left(\begin{array}{cccccccc}
		0 & 1 &2 &3 &4 &5 &6 &7 \\
        1 & 0 &3 &2 &5 &4 &7 &6 \\
        2 & 3 &0 &1 &6 &7 &4 &5 \\
        3 & 2 &1 &0 &7 &6 &5 &4 \\
        4 & 6 &5 &7 &0 &2 &1 &3 \\
        5 & 7 &4 &6 &1 &3 &0 &2 \\
        6 & 4 &7 &5 &2 &0 &3 &1 \\
        7 & 5 &6 &4 &3 &1 &2 &0 
	\end{array}\right)\]
	Because $F$ is a finate set, $g_i$ is a injection and invertible. For any operator $g_i$, it exist an inverse operator and we denote it as $g^{-}_{i}$. Operator $g_i^-$ is also a surjection and injection in $F$, which can be simply explained by the following matrix $A^-$. In matrix $A^-$, $A^-_{ij} = k$ means $g_i^-(f_j) = f_k$. For simplification, we omit $f$ and use $g_i^-(j) = k$ to express the results above.
		\[A^- = \left(\begin{array}{cccccccc}
		0 & 1 &2 &3 &4 &5 &6 &7 \\
        1 & 0 &3 &2 &5 &4 &7 &6 \\
        2 & 3 &0 &1 &6 &7 &4 &5 \\
        3 & 2 &1 &0 &7 &6 &5 &4 \\
        4 & 6 &5 &7 &0 &2 &1 &3 \\
        6 & 4 &7 &5 &2 &0 &3 &1 \\
        5 & 7 &4 &6 &1 &3 &0 &2 \\
        7 & 5 &6 &4 &3 &1 &2 &0 
	\end{array}\right)\]
	
	Based on the above premise, we built the predicting method by the follow 4 steps. Figure \ref{fig:1} shows the method by a flow chart.
	\begin{enumerate}
		\item Apply $(f_0,f_1\dots ,f_7)$ to the image $X_t$ to get 8 images. We denote these images as $(X_{t0},X_{t1},\dots,X_{t7}) $
		\item Take these 8 images as input to DNN and get 8 orientations $(i_{t0},i_{t1},\dots,i_{t7})$.
		\item Apply $(g^-_{0},g^-_{1},\dots,g^-_{7})$ to these 8 labels and get another 8 labels $(g^-_{0}(i_{t_0}),g^-_{1}(i_{t1}),\dots,g^-_{7}(i_{t7}))$. 
		\item The labels which occur most in  $(g^-_{1}(i_{t_0}),g^-_{2}(i_{t1}),\dots,g^-_{7}(i_{t7}))$ is the final result.
	\end{enumerate}
\section{Experiment}
\subsection{Experiment Setup}
\begin{figure}[htb]
\centering 
\subfloat[C0]{

  \includegraphics[height=0.2\linewidth]{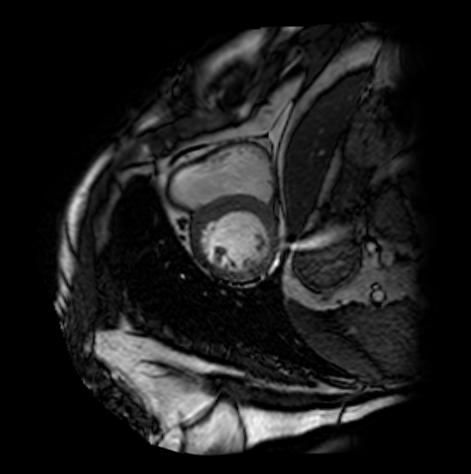}
}
\hspace{10pt}
\subfloat[LGE]{ 

	\includegraphics[height=0.2\linewidth]{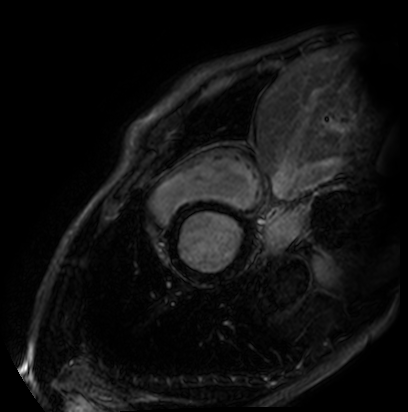}
} 
\hspace{10pt}
\subfloat[T2]{ 

	\includegraphics[height=0.2\linewidth]{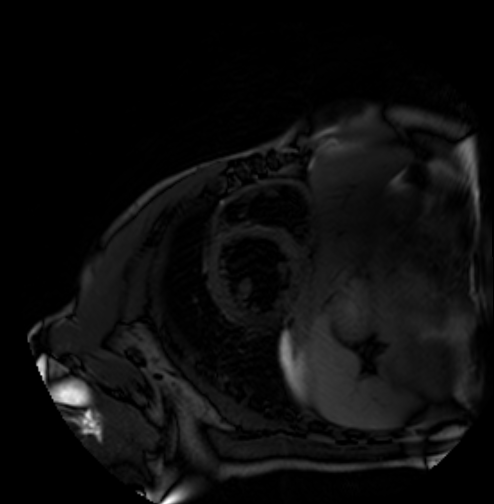}
} 
\caption{the three-sequence Cardiac Magnetic Resonance} 
\label{fig:2}
\end{figure}

We evaluate orientation recognition network on the MyoPS dataset\cite{R:1,R:2}. The MyoPS dataset provides the three-sequence Cardiac Magnetic Resonance (LGE, T2 and C0) and three anatomy masks, including myocardium (Myo), left ventricle (LV), and right ventricle (RV), some of the three-sequence Cardiac Magnetic Resonance is shown as Figure \ref{fig:2} . MyoPS further provides two pathology masks (myocardial infarct and edema) from the 45 patients. For the simplified orientation recognition network, we train model for single modality on the MyoPS dataset, then transfer the model to other modalities. For each sequence, we resample each slice of each 3d image and the corresponding labels to an in-plane resolution of 1.367 × 1.367 mm.

We divide slices into three sub-sets, i.e., the training set, validation set and test set, at the ratio of 50\%, 30\% and 20\%. Three sub-sets don't have slices from same patient. Then, for each standard 2d image, we apply all function from $F$ to it to expand dataset. For training set, image slices are cropped or padded to $256 \times 256$ for the orientation recognition network, and apply random augmentation. For test set and validation set, the images are only resized to $256 \times 256$.
\subsection{Orientation Recognition Network}
During each training iteration, a batch of the three-channel images X' is fed into the orientation recognition network. Then, the network outputs the predicted orientation network. 

\begin{figure}[htb]
\centering 
\subfloat[C0]{

  \includegraphics[height=0.2\linewidth]{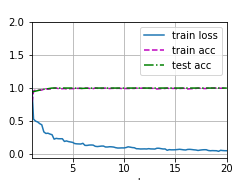}
}
\hspace{10pt}
\subfloat[LGE]{ 

	\includegraphics[height=0.2\linewidth]{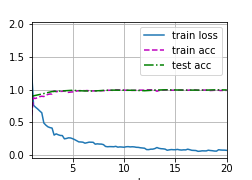}
} 
\hspace{10pt}
\subfloat[T2]{ 

	\includegraphics[height=0.2\linewidth]{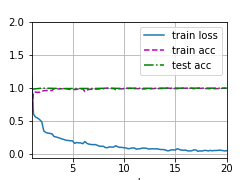}
} 
\caption{Training images} 
\label{fig:3}
\end{figure}

Figure \ref{fig:3} and Table \ref{tab:2} shows the training process and accuracy of the three sequences on test set. The results show that the model get quite high accuracy in three modalities.  Howeverm the size of test data is small, random factors influence the result heavily. In the following, we redivide the data,   retrain the model, and analyses sensitivity of the model.

\begin{table}[htb]
\centering
\caption{Accuracy of the three Modalities}
\begin{tabular}{cccc}
\toprule
 Modality & Accuracy & Description \\
 \midrule
C0 & 1.0 & Pre-train\\
LGE & 1.0  & Transfer learning\\
T2 & 1.0  & Transfer learning\\
 \bottomrule
\end{tabular}
\label{tab:2}
\end{table}

\subsection{Sensitivity analysis}
	When using deep learning method to solve the problem, the volume of data is always the most important factor. In medical image processing, it is difficult to get a lot of data, so analysing the sensitivity of model is necessary. We redivided the data set into training set and test set 5 times, and the proportion of training set was 60\%, 50\%, 40\%, 30\% and 20\% respectively. In each divided data set, we retrain the model in training set and compute accuracy in test set.
	
	The accuracy variation is shown in Table \ref{tab:3}. There is not a distinct difference among the accuracy while the volume of data decrease. Therefore DNN is a suitable method in CMR orientation recognition with high accuracy and low sensitivity.
	
\begin{table}[htb]
\centering
\caption{Accuracy of different training data size, modalities and prediction methods. The left column is the ratio of training data. Improved prediction means the method we proposed in 2.3 while Direct prediction means inputing the resized image to the DNN directly}
\label{tab:3}
\begin{tabular}{cccc|ccc}
 \toprule
 & \multicolumn{3}{c}{Improved prediction}  & \multicolumn{3}{c}{Direct prediction}\\
 \cmidrule{2-4}  \cmidrule{5-7} 
 & C0 & LGE & T2 & C0 & LGE & T2\\
 \midrule
60\% &1.0 & 1.0 & 1.0 & 1.0 & 1.0 & 0.998\\
50\% &1.0& 1.0& 1.0 & 1.0 & 0.991 & 0.999\\
40\% &0.974 &1.0&1.0 & 0.967& 0.996 & 0.996\\
30\% &1.0&0.979&1.0 & 1.0 &0.975 & 0.997\\
20\% &0.982 &0.960&0.986 & 0.978& 0.953 & 0.983\\
 \bottomrule
 \end{tabular}

\end{table}
\subsection{Comparison between improved prediction and direct prediction}
	Table \ref{tab:3} shows the difference of accuracy  between improved prediction and direct prediction. We can find that improved prediction always have higher accuracy in our experiment. However, it is not inevitable, because the vote may make the original correct decision wrong. Sometimes, the improved prediction have a lower accuracy, but in an average sense, the improved prediction is better than direct prediction.

\section{Conclusion}
DNN model get quite high accuracy in recognition of CMR image orientation and transfer learning make it easy to be transferred to other modalities . Thanks to the data expansion and augmentation, the model only need a few data. The improved prediction we proposed further increase the accuracy. We are sure that DNN model combining with transfer learning and improved prediction can be used in other recognition of orientation tasks.

\bibliography{main}

\begin{thebibliography}{1}

\bibitem{R:3}
Ke~Zhang and Xiahai Zhuang.
\newblock Recognition and {Standardization} of {Cardiac} {MRI} {Orientation}
  via {Multi}-tasking {Learning} and {Deep} {Neural} {Networks}.
\newblock In Xiahai Zhuang and Lei Li, editors, {\em Myocardial {Pathology}
  {Segmentation} {Combining} {Multi}-{Sequence} {Cardiac} {Magnetic}
  {Resonance} {Images}}, Lecture {Notes} in {Computer} {Science}, pages
  167--176, Cham, 2020. Springer International Publishing.

\bibitem{R:2}
Xiahai Zhuang.
\newblock Multivariate mixture model for cardiac segmentation from
  multi-sequence mri.
\newblock In {\em International Conference on Medical Image Computing and
  Computer-Assisted Intervention}, pages 581--588. Springer, 2016.

\bibitem{R:1}
Xiahai Zhuang.
\newblock Multivariate mixture model for myocardial segmentation combining
  multi-source images.
\newblock {\em IEEE transactions on pattern analysis and machine intelligence},
  41(12):2933--2946, 2019.

\end{thebibliography}

\end{document}